\documentclass{article}

\usepackage{PRIMEarxiv}
\usepackage[utf8]{inputenc} 
\usepackage[T1]{fontenc}    
\usepackage{hyperref}       
\usepackage{url}            
\usepackage{booktabs}       
\usepackage{amsfonts}       
\usepackage{nicefrac}       
\usepackage{microtype}      
\usepackage{lipsum}
\usepackage{fancyhdr}       
\usepackage{graphicx}       
\usepackage{bm}
\graphicspath{{media/}}

\newcommand{\tabincell}[2]{\begin{tabular}{@{}#1@{}}#2\end{tabular}}

\title{A Lightweight Dual-Domain Attention Framework for Sparse-View CT Reconstruction
}

\author{
  Chang Sun, Ken Deng, Yitong Liu, Hongwen Yang \\
  Beijing University of Posts and Telecommunications \\
  Beijing, China\\
  \texttt{\{sc1998, arieldeng, liuyitong, yanghong\}@bupt.edu.cn} \\
}

\begin{document}
\maketitle

\begin{abstract}
Computed Tomography (CT) plays an essential role in clinical diagnosis. Due to the adverse effects of radiation on patients, the radiation dose is expected to be reduced as low as possible. Sparse sampling is an effective way, but it will lead to severe artifacts on the reconstructed CT image, thus sparse-view CT image reconstruction has been a prevailing and challenging research area. With the popularity of mobile devices, the requirements for lightweight and real-time networks are increasing rapidly. In this paper, we design a novel lightweight network called CAGAN, and propose a dual-domain reconstruction pipeline for parallel beam sparse-view CT. CAGAN is an adversarial auto-encoder, combining the Coordinate Attention unit, which preserves the spatial information of features. Also, the application of Shuffle Blocks reduces the parameters by a quarter without sacrificing its performance. In the Radon domain, the CAGAN learns the mapping between the interpolated data and fringe-free projection data. After the restored Radon data is reconstructed to an image, the image is sent into the second CAGAN trained for recovering the details, so that a high-quality image is obtained. Experiments indicate that the CAGAN strikes an excellent balance between model complexity and performance, and our pipeline outperforms the DD-Net and the DuDoNet.
\end{abstract}


\section{Introduction}
In recent years, the significance of Computed Tomography (CT) in clinical diagnosis has seen rapid growth, especially under the background that the COVID-19 is globally spreading \cite{chung2020ct}. Doctors need CT for diagnosis not only in sophisticated hospitals equipped with workstations, but also in rural clinics that cannot afford too much computing power. Therefore, fast, lightweight, and high-quality CT imaging becomes crucial. Since the radiation will lead to the risk of inducing cancer, scholars have been dedicated to reducing the radiation dose as low as reasonably achievable (ALARA) \cite{slovis2002the}. Sparse sampling is a prevailing method of radiation dose reduction, which may cause severe streak artifacts in the images. Hence the trade-off between the sparsity of projection views and image quality is worthy of study and has raised much attention. In this work, we focus on the strategy of obtaining high-quality images from sparse-view projection data.

\begin{figure}[tbhp]
\centerline{\includegraphics[width=14cm]{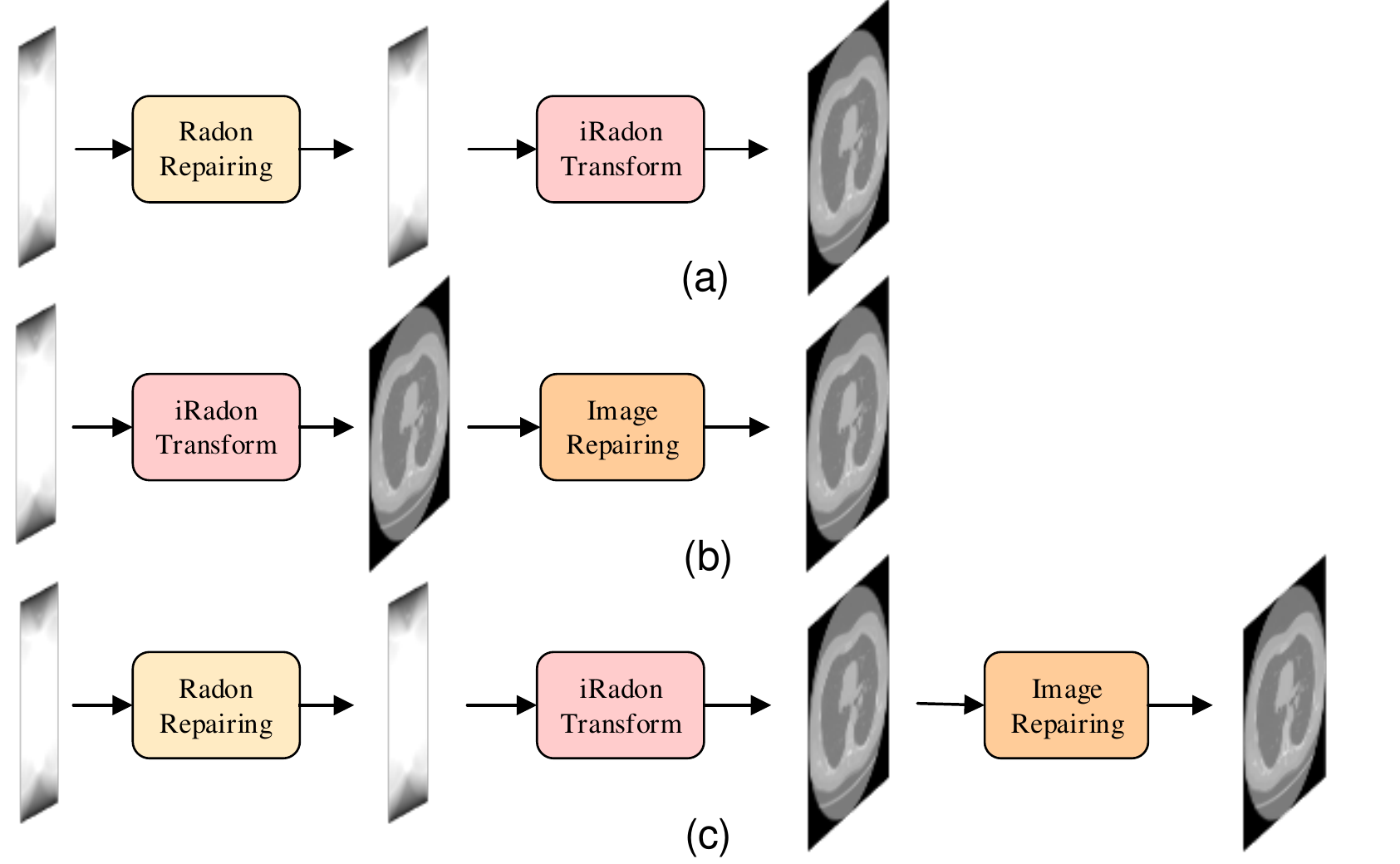}}
\caption{Three mainstream sparse-view CT reconstruction pipelines. The inputs are sparse-view CT projection data. (a) Radon complementing. (b) Image post-processing. (c) Dual-domain restoration.}
\label{sec:fig0}
\end{figure}

The traditional analytical CT reconstruction method FBP \cite{katsevich2002theoretically} requires sufficient sampling, and the reconstruction becomes a challenging inverse problem when the scanning angles are restricted. In recent years, iterative algorithms \cite{Andersen1984Simultaneous, Hu2017An, chen2018learn, Zhang2019JSR} have been dominant methods showing their great effectiveness, but they are usually time-consuming, highly relying on human experience and parameter fine turning.

In the past few years, deep learning methods based on Convolutional Neural Networks (CNN) have been adopted extensively to different kinds of image problems. With the Generative Adversarial Networks (GAN) \cite{goodfellow2014generative} developing rapidly, it has been used extensively to generate vivid images. Therefore, deep learning-based CT reconstruction approaches have been developed, and networks based on GANs have become nonnegligible.

Generally, sparse-view CT image reconstruction methods based on deep learning can be grouped into three categories: Radon complementing, image post-processing, and dual-domain restoration, as shown in Figure \ref{sec:fig0}.

The incompletion of projection views may cause severe streak artifacts on reconstructed images. Therefore, sinogram complementing is useful for streak artifact removal, which has become a prevailing method since 2018. Generally, it first conduct complementing on the sinogram, then reconstruct it into image \cite{Bai2018Limited, Dong2019A, fu2020a, Anirudh2018Lose, Dai2019Limited, ghani2018deep, lee2019deep, dong2019sinogram}. However, the information cannot remain intact through CT reconstruction algorithms, and the image quality depends on the number of projection views to a great extent. Besides, the complementing ability in the Radon domain is limited. Therefore, there are still some blur and losses of details in the reconstructed images after sinogram complementing.

Image post-processing is to reconstruct the sparse Radon data to images and implement post-processing on images. Researchers build different models to learn the mapping between the images with artifacts and without artifacts \cite{zhang2016image, Zhang2018A, kuanar2019low, guan2020fully}. In general, there remains difficulty to remove severe streak artifacts completely, since the distribution of the streak artifacts is difficult to be modeled.

To remedy the insufficient accuracy of methods in a single domain, researchers have proposed dual-domain methods to simultaneously utilize the information both in the Radon domain and the image domain since 2018 \cite{hammernik2017a, liang2018comparison, Zhao2018Sparse, Lee2018High, lin2019dudonet, yin2019domain, Zhu2020Low, Zhang2020Artifact, hu2021hybrid, zhou2022dudodr}. These works have proved that dual-domain methods can make full use of information both in projection and image domains, where the information in dual domains is supplementary to each other, therefore dual-domain methods perform better than single-domain methods \cite{hu2021hybrid}.

In this work, we designed a lightweight dual-domain sparse-view CT reconstruction pipeline and established a novel adversarial network called CAGAN. In the Radon domain, the generator of CAGAN learns the mapping between the interpolated and fringe-free projection data. After the restored projection data is reconstructed to an image, the image is sent into the second CAGAN trained for recovering the textures, so that a clear and high-quality image is obtained.

Our contributions can be summarized as follows: (1) We designed a lightweight U-shape adversarial network called CAGAN with a modified version of Shuffle block, keeping the performance while reducing the parameters to a quarter. (2) We fuse the coordinate attention module in our network, which learns to preserve the positional information and focus on the textures. (3) We adopt CAGAN to implement a dual-domain end-to-end sparse-view CT reconstruction framework and obtain clear and detailed CT images.

\section{Related Works}

\textbf{Lightweight architectures. } With the popularity of mobile devices growing in recent years, it is obliged for deep neural networks to meet the needs of fast speed and low complexity without sacrificing performance, which depends on the efficiency of basic module designs to a great extent. Several attempts have been made to realize lightweight models. For instance, MobileNets \cite{howard2017mobilenets, Sandler2018MobileNetV2, howard2019searching} utilized depthwise and pointwise convolutions to construct a unit for approximating the original convolutional layer with larger filters, achieving comparable performance. ShuffleNets \cite{zhang2018shufflenet, ma2018shufflenet} further proposed a channel shuffle operation to make the features in different channels communicate with each other and enhance the performance of lightweight models. They are not only beneficial for the network performance, but also plug-and-play and convenience to be embedded into the previous networks. 

\textbf{Attention. }Recent researches have shown that the representations produced by CNNs can be strengthened by integrating learning mechanisms that help to capture spatial \cite{jaderberg2015spatial} or channel-wise \cite{hu2018squeeze} correlations between features. Among those works, the “Squeeze-and-Excitation” (SE) structure \cite{hu2018squeeze} exhibits its extraordinary capability, which adaptively recalibrates channel-wise feature responses by explicitly modeling interdependencies between channels. However, SE-Net will sacrifice the spatial information in the feature maps. The latest proposed Coordinate Attention (CA) \cite{hou2021coordinate} makes some progress aiming at the disadvantage of SE-Net. It factorizes channel attention into two 1D feature encoding processes that aggregate features along the two spatial directions, respectively. In this way, long-range dependencies can be captured along one spatial direction, and meanwhile precise positional information can be preserved along the other spatial direction. Besides, embedding CA in the network only leads to a few extra parameters, which is suitable to be deployed to lightweight networks. 

\section{Methods}

\begin{figure*}[ht]
\centerline{\includegraphics[width=\linewidth]{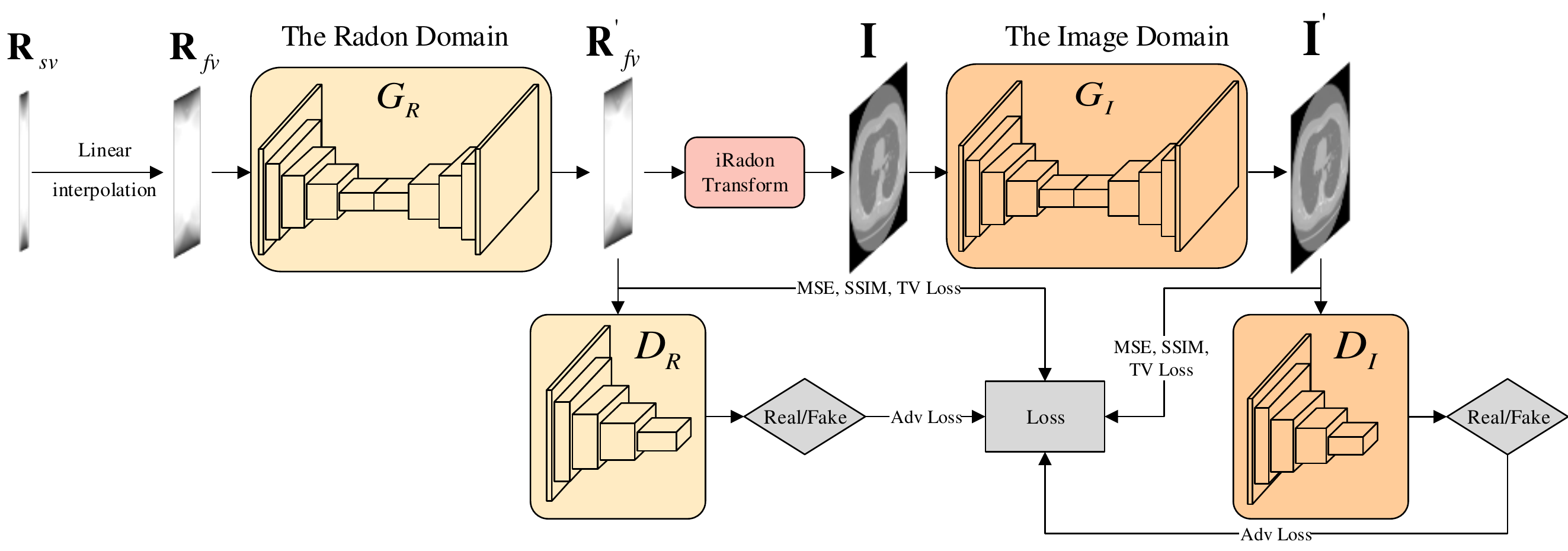}}
\caption{The overall structure of our proposed pipeline.}
\label{sec:fig1}
\end{figure*}

\subsection{Pipeline Overview}

The universal approximation theorem \cite{hornik1989multilayer} points out that multilayer feedforward networks can approximate various continuous functions. Thus, deep neural networks are suitable to fit the restoration functions both in the Radon domain and image domain. 

In this paper, we consider data with 180 views as full-view data. As for the data preprocessing, the sparsely sampled Radon data $\mathbf{R}_{sv}$ is bilinearly interpolated into 180 views, obtaining $\mathbf{R}_{fv}$. Interpolation is the inverse process of sparsely sampling, which can roughly fill the missing projection views and lower the difficulty of learning for the network.

In stage one, $\mathbf{R}_{fv}$ is sent into the CAGAN of the Radon domain for restoration, whose output is Radon data ${\mathbf{R}_{fv}}^\prime$ with 180 views. The above process can be represented as ${\mathbf{R}_{fv}}^\prime=f(\mathbf{R}_{fv})$, where function $f$ is fitted by network $G_R$. Afterwards, ${\mathbf{R}_{fv}}^\prime$ is reconstructed into CT image $\mathbf{I}$ by FBP algorithm \cite{katsevich2002theoretically}. The complementing process in the Radon domain can eliminate jagged edges in the interpolated Radon data and solve the blurring and angular deflection in the reconstructed images, increasing the quality of reconstructed images but remaining the problem of lack of textures and details. 

To acquire high-quality CT images, we implement post-processing on the reconstructed image $\mathbf{I}$ as stage two. As a result, clear and detailed images are generated, which process can be represented as $\mathbf{I} =g(\mathbf{I})$, where function $g$ is approximated by network $G_I$. 

In both the Radon domain and the image domain, we use the generator of the CAGAN to fit functions $f$ and $g$. Despite the same architecture, the tasks in two stages are different and distinguishable. Network $G_R$ learns to correct the angular deflection in the interpolated projection data in the Radon domain, while network $G_I$ manages to recover the missing details in the reconstructed images. Finally, we manage to obtain ideal results through the dual-domain CT reconstruction pipeline. Figure \ref{sec:fig1} shows the overall pipeline of our method.  

\subsection{CAGAN}

\begin{figure}[bhtp]
\centerline{\includegraphics[width=\linewidth]{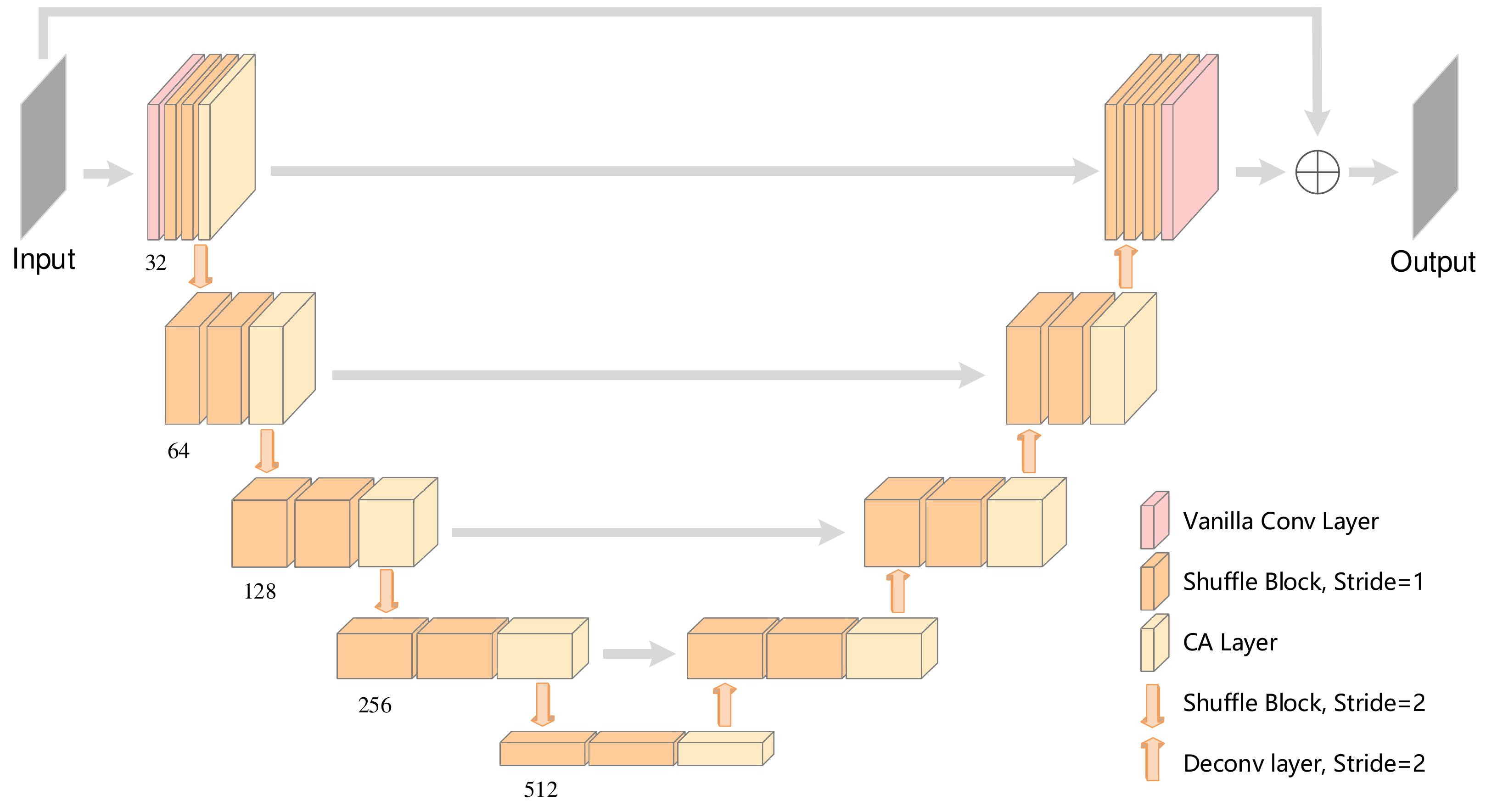}}
\caption{The structure of the generator of CAGAN. }
\label{sec:fig2}
\end{figure}

We designed CAGAN, which is an adversarial auto-encoder with lightweight convolutional unit Shuffle Blocks and Coordinate Attention Blocks \cite{hou2021coordinate}. The generator of CAGAN is a U-shape network, and experiments have proved that auto-encoders in U-shape are remarkably useful for image enhancement and inpainting \cite{liang2018comparison, lee2019deep}. The corresponding architecture of the generator of CAGAN is illustrated in Figure \ref{sec:fig2}.

In the encoder, we initially adopt a convolutional layer, then we perform the pattern of "2 Shuffle Blocks + 1 CA Block + Down sampling" for 4 times. At the bottom of the "U", we obtain the encoded semantic features. As for the decoder, we use the symmetrical structure of "Deconvolutional layer + 2 Shuffle Blocks + 1 CA Block" for 4 times. Shortcut connections are applied to concatenate the features in the contracting path to the corresponding features in the expansive path, which is conducive to the preservation of the spatial structures. In the end, a $1\times1$ convolution is carried out to fuse the feature maps into a one-channel picture.

The structure of the discriminator is similar to the structure of the encoder in the generator. The discriminator can extract the high-level semantic features of an image similarly to the encoder, and outputs the possibility that an image is real through a sigmoid function. During the training process, the discriminator learns to distinguish between real images and fake images generated by the generator. The possibility is treated as a term in the loss function, which will conduce to generating clearer images for the generator. 

\subsubsection{Shuffle Block}

Shuffle Block used the considerable experience of ShuffleNet V2 \cite{ma2018shufflenet}. It adopts depth separable convolution, which could be divided into depthwise convolution and pointwise convolution, and such operations can reduce the number of parameters without suffering from performance deterioration.

\begin{figure}[htbp]
\centerline{\includegraphics[width=\linewidth]{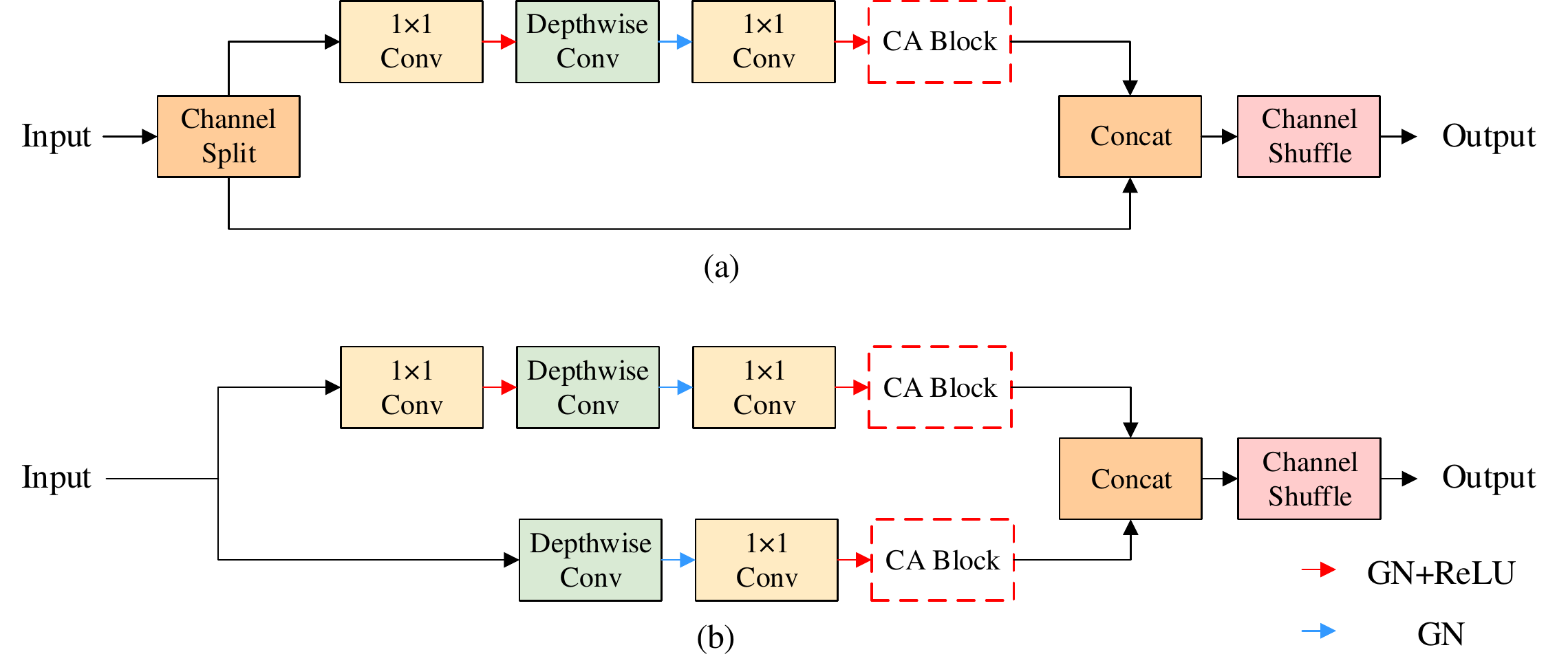}}
\caption{The structure of Shuffle Block. The version with CA Block is referred to as "CA inside", and the CA Blocks are removed and placed after the Shuffle Blocks in the final version of CAGAN.}
\label{sec:fig3}
\end{figure}

The structure of Shuffle Block is shown in Figure \ref{sec:fig3}. We modified the batch normalization (BN) \cite{ioffe2015batch} in ShuffleNet V2 into group normalization (GN) \cite{wu2018group}, for the reason that the batchsize is limited to 2 under the restriction of memory. Researches have shown that GN may illustrate better performance than BN when batch size is smaller than 8 \cite{wu2018group}.

When we don't want the size of feature maps to change, $stride$ is 1 and the number of input channels equals the number of output channels, the input is equally split into two branches, as shown in Figure \ref{sec:fig3} (a). One branch remains identity, while the other branch consists of three convolution layers. In the end, the two branches are concatenated and shuffled by channel to enable information communication between the two branches. "Channel shuffle" is often used after group convolution, since group convolution can reduce the calculation but also have a side effect that outputs from a certain channel are only derived from a small fraction of input channels, and "channel shuffle" can solve it \cite{zhang2018shufflenet}. Otherwise, when we want use the shuffle block as a down-sampling operator, the channel split operator is removed, so that the number of output channels can be different from the number of input channels, whose architecture is illustrated in (b).

Also, we found that adopting Shuffle Block with a stride of 2 to perform downsampling instead of max-pooling achieves better performance, because convolutional blocks can extract features more precisely and keep more information. 

\subsubsection{Coordinate Attention}

In the image domain, the amount of information in different regions varied wildly because of the specialty of CT images. Specifically, the texture details in the center of the CT images are the critical point in the diagnosis, while the background and the surroundings of the image are pure black or grey without carrying any information. Thus, we hope to use the attention mechanism, leading the network to focus on the texture details and obtain higher quality images \cite{wang2020improving}. To keep the model lightweight, we need an attention mechanism which is not too computation-consuming, but capable of learning the interested areas in the feature maps.

\begin{figure}[htbp]
\centerline{\includegraphics[width=\linewidth]{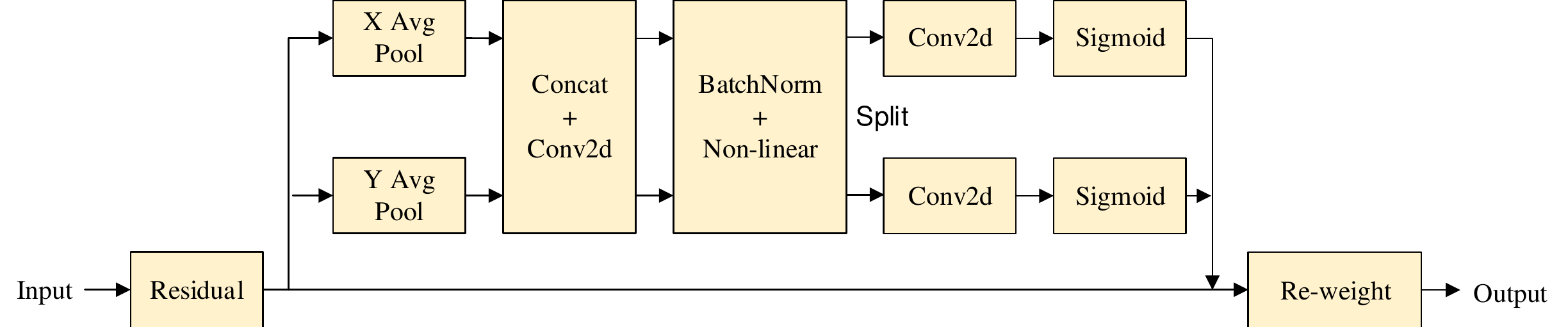}}
\caption{The structure of CA Block.}
\label{sec:figca}
\end{figure}

Coordinate attention \cite{hou2021coordinate} is a perfect choice, which factorizes channel attention into two 1D feature encoding processes that aggregate features along horizontal and vertical directions. The two attention maps can be complementarily applied to the input feature map, augmenting the representations of the interested area. This process allows CA to locate the exact position of the object of interest more accurately and hence helps the model predict more precisely. The extra computational cost is nearly neglectable, and the model complexity can be adjusted by the hyper-parameter $r$ in the non-linear layer of CA, as is shown in Figure \ref{sec:figca}.

\subsubsection{Loss}
When training the pipeline end-to-end, the total loss is defined as equation (\ref{eq0}), and illustrated in Figure \ref{sec:fig1}.

\begin{equation}
    l_{total}=l_{G_{R}}+l_{G_I}
    \label{eq0}
\end{equation}

On the single domain, we deployed the loss function for the generator as equation (\ref{eq2}), which consists of four terms $l_{MSE}$, $l_{SSIM}$, $l_{adv}$ and $l_{tv}$, with their respective hyperparameters $\alpha_1$, $\alpha_2$, $\alpha_3$ and $\alpha_4$. According to the empirical values in \cite{ledig2017photo, liu2018image}, they are set to 1, 1, 1e-3, and 1e-1.

\begin{equation}
    l_{G}=\alpha_{1}l_{MSE}+\alpha_{2}l_{SSIM}+\alpha_{3}l_{adv}+\alpha_{4}l_{tv}
    \label{eq2}
\end{equation}

$l_{MSE}$ calculates the L2 loss. We take the Radon domain as an example, ${\mathbf{R}_{fv}}^\prime$ and $\mathbf{R}_{gt}$ stand for the output of Radon domain and its corresponding ground truth, and $l_{MSE}$ is represented as $l_{MSE}={\Vert \mathbf{R}_ {gt} -  {\mathbf{R}_{fv}}^\prime \Vert }_2$. 

However, pixel-wise evaluation such as MSE cannot completely measure the quality of an image, because human visual perception is highly adapted for extracting structural information from a scene. As a result, SSIM (Structural Similarity) \cite{wang2004image} was proposed, which measures the structural similarity of two pictures from three aspects of luminance, contrast, and structure. An SSIM closer to 1 demonstrates that the two images possess a higher structural similarity. Hence $l_{SSIM}$ is defined as $l_{SSIM} = 1-SSIM \left( \mathbf{R}_{fv} ^\prime ,\mathbf{R}_{gt} \right)$.

$l_{adv}$ refers to the adversarial loss, and it is defined as $l_{adv}=1-D\left(\mathbf{R}_{fv}^\prime\right)$, where $D\left(\mathbf{R}_{fv}^\prime\right)$ means the probability that the fake image $\mathbf{R}^\prime_{fv}$ is real. Therefore, $l_{adv}$ punishes the probability that the fake image generated by $G$ is identified as a false image.

To encourage spatial smoothness in the output image, we follow the prior works in image denoising \cite{rudin1992nonlinear} and super resolution \cite{aly2005image}, and compute total variation (TV) regularization term $l_{tv}\left(\mathbf{R}_{fv}^\prime\right)$.

For the discriminator, the loss function is shown in equation (\ref{eq7}).

\begin{equation}
    l_{D}=1-D\left(\mathbf{R}_{gt}\right)+D\left(\mathbf{R}_{fv}^\prime\right)
    \label{eq7}
\end{equation}
	
\subsection{Network training}

During the training process, we adopt the Adam algorithm for optimization. In the Radon domain, we trained the CAGAN for 20 epochs, and the learning rate is set to 3e-4 initially, and decreased down to 3e-5 after 10 epochs. In the image domain, we trained the network for 30 epochs, and the learning rate is set to 1e-4 initially and decreased down to 1e-5 for the last 20 epochs. Finally, we combined the dual-domain end-to-end pipeline and finetuned it for 30 epochs. Besides, we implement our structure using PyTorch \cite{paszke2019pytorch} on a GeForce RTX 2080 Ti. 

\section{Experiments and Results}

\subsection{Dataset and evaluation}
We adopt LIDC-IDRI \cite{armato2011the} as our dataset, which includes 1018 cases and about 240 thousand CT images with a size of $512\times 512$. The slice thickness of CT images varies from 0.6 mm to 5.0 mm with a median of 2.0 mm. Case 1-100 is used as the test set, case 101-200 is used as the validation set, and the rest is the training set. The origin CT image is transformed into parallel beam CT projection data of 180 projection views, as the ground truth in the Radon domain $\mathbf{R}_{gt}$. It is uniformly sampled into 45, 20, 10 projection views respectively. We simulate poisson noise on the sparse-view Radon data, assuming the incident X-ray has $2 \times 10^7$ photons. The sparse-view data is referred to as $\mathbf{R}_{sv}$. We interpolated it into a matrix of $512 \times 180$, which is the input of the model $\mathbf{R}_{fv}$. Before being sent into the model, it is padded into a matrix of $512 \times 192$ to fit the model, and we remove the padding in the testing procedure. 

We choose PSNR (Peak Signal to Noise Ratio) and SSIM \cite{wang2004image} as the evaluation metrics. We will show some ablation studies to analyze each part of our network in the case of 45 projection views mostly in the Radon domain. The data shown in the following tables are mean values computed on the test set.

\subsection{Ablation Study}

\subsubsection{The performance of The Shuffle Block}

To verify the performance of Shuffle Blocks, we replace all of them with vanilla convolutional blocks in the model (including the downsampling operators), which are in the manner of "conv2d + BN + ReLU". The kernel size of conv2d is $3 \times 3$, and other hyperparameters remain the same, with a $r$ of 16 and a $g$ of 32. The result in the Radon domain is presented in the second row and the last row in Table \ref{table1}. The $parameters$ and $GFLOPs$ stand for the number of parameters and the number of floating-point operations in the generator.

\begin{table}[htbp]
  \centering
    \begin{tabular}{ccccc}
    \hline
    {} &
    \tabincell{c}{Parame\\ters(M) } &
    GFLOPs &
    \tabincell{c}{Radon\\PSNR/SSIM} &
    \tabincell{c}{Image\\PSNR/SSIM} \\
    \hline
    (a) & -     & -     & 45.271/0.989 & 28.722/0.833 \\
    (b) & 8.641 & 19.18 & \textbf{59.275/0.9989} & \textbf{30.966/0.920} \\
    (c) & 1.837 & 3.69  & 57.590/0.9985 & 30.850/0.916 \\
    \textbf{(d)} & \textbf{2.026} & \textbf{4.23} & 58.849/0.9988 & 30.929/0.917 \\
    \hline
    \end{tabular}%
    \caption{The comparison of Shuffle Block with Vanilla Convolutional Block and max-pooling. (a) The original input. (b) Vanilla Conv Blocks. (c) Maxpooling. (d) Shuffle Blocks.}
  \label{table1}%
\end{table}%

As shown in Table \ref{table1}, the parameters and flops can be reduced by more than 4 times using Shuffle Blocks, with only slight loss of the performance. The result fully reflects the lightweight characteristics of the Shuffle Block so that it can be easily deployed on mobile and miniaturized CT equipment.

Besides, we replace all of the downsampling Shuffle Blocks with max-pooling operators, as shown in the third row of Table \ref{table1}. The result indicates that the performance of Shuffle Blocks is more superior, which may be due to the difference that convolution is more powerful for feature extraction.

\subsubsection{The significance of the discriminator}
To investigate the significance of the discriminator in CAGAN, we removed it and the corresponding term of $l_{adv}$, lefting the generator alone to learn to generate restored images. It can be seen from Table \ref{tab:table3} that adding the discriminator is conducive to the improvement of PSNR and SSIM of the restored image.

\begin{table}[htbp]
  \centering

    \begin{tabular}{ccc}
    \hline
    {} & \tabincell{c}{Radon \\ PSNR/SSIM} & \tabincell{c}{Image \\ PSNR/SSIM} \\
    \hline
    \textbf{With Discriminator} & \textbf{58.849/0.9988} & \textbf{30.929/0.917} \\
    Without Discriminator & 58.010/0.9987 & 30.910/0.916 \\
    \hline
    \end{tabular}%
    \caption{The comparison between with and w/o discriminator.}
  \label{tab:table3}%
\end{table}%

\subsubsection{The effect and position of CA}

In this part, we explored the answers to the following questions: Is CA efficient? What is the difference between CA and the prevailing SE? What is the best position of CA? Therefore, we carried out four experiments as shown in Table \ref{table2}: (1) remove all the CA blocks in CAGAN; (2) put CA block inside each of the Shuffle Blocks, as illustrated in Figure \ref{sec:fig3}; (3) place the CA block outside after each two Shuffle Blocks; (4) put SE blocks after each two Shuffle Blocks.

It is indicated that CA promotes the restoration performance in both the two stages no matter the position, compared with the network without attention, while the number of extra parameters is only $4\%$ of the origin. Further, placing CA blocks after each two Shuffle Blocks can achieve a better effect. However, SE can only outperform the network without attention in the stage one.

\begin{table*}[htbp]
  \centering
    \begin{tabular}{cccccc}
    \hline
    {} &
    Parameters(M) &
    GFLOPs &
    \tabincell{c}{Radon PSNR/SSIM \\ (Stage1)} &
    \tabincell{c}{Image PSNR/SSIM \\ (Stage1)} &
    \tabincell{c}{Image PSNR/SSIM \\ (Stage2)} \\
    \hline
    Without attention & 1.944 & 4.23  & 57.471/0.9987 & 30.916/0.917 & 41.315/0.962 \\
    CA inside & 2.014 & 4.23  & 58.445/0.9988 & 30.921/0.916 &  {41.634/0.963} \\
    \textbf{CA outside} & 2.026 & 4.23  & \textbf{58.849/0.9988} & \textbf{30.929/0.917} & \textbf{41.975/0.963} \\
    SE outside & 1.999 & 4.25  & 58.028/0.9986 & 30.897/0.917 & 41.082/0.959 \\
    \hline
    \end{tabular}%
    \caption{Restoration Results by different kinds and postions of attention module.}
  \label{table2}%
\end{table*}%

\subsubsection{BN VS. GN}

Most researchers habitually use batch normalization \cite{ioffe2015batch} now. Nonetheless, theoretical studies have shown that when the batch size is small, group normalization is more effective than batch normalization. Therefore, we replace BN in Shuffle Blocks with GN, where the batchsize is 1, and other parameters keep the same. As shown in Table \ref{tab:table4}, GN is more conducive to the improvement of model performance than BN.

\begin{table}[htbp]
  \centering
    \begin{tabular}{ccc}
    \hline
    {}  & \tabincell{c}{Radon \\ PSNR/SSIM} & \tabincell{c}{Image \\ PSNR/SSIM}\\
    \hline
    \textbf{Group Normalization} & \textbf{58.849/0.9988} & \textbf{30.929/0.917} \\
    Batch Normalization & 58.341/0.9987 & 30.926/0.916 \\
    \hline
    \end{tabular}%
    \caption{The comparison of BN and GN.}
  \label{tab:table4}%
\end{table}%

\subsubsection{The selection of r in CA}

There are $C/r$ input neurons and $C$ output neurons in the non-linear layer of the CA block, which means the hyperparameter $r$ controls the channel shrinkage ratio and the number of parameters in the CA block. To find out the optimal solution of $r$ in the CA Block, we carried out three experiments. Since the least number of the channels of the features in our network is 32, $r$ is set to 16, 8, and 4 respectively. Other hyperparameters keep the same. As shown in Table \ref{tab:table5}, our results are the same as the conclusion in \cite{hou2021coordinate}: 16 is the optimal value of $r$, in terms of both parameters and performance.

\begin{table}[htbp]
  \centering
    \begin{tabular}{ccccc}
    \hline
    $r$     &
    \tabincell{c}{Parame\\ ters(M)} &
    GFLOPs &
    \tabincell{c}{Radon\\PSNR/SSIM} &
    \tabincell{c}{Image\\PSNR/SSIM} \\
    \hline
    \bm{$16$} & \textbf{2.026} & \textbf{4.23} & \textbf{58.849/0.9988} & \textbf{30.929/0.917} \\
    $8$   & 2.108 & 4.24  & 58.458/0.9988 & 30.907/0.916 \\
    $4$   & 2.272 & 4.25  & 58.102/0.9987 & 30.894/0.914 \\
    \hline
    \end{tabular}%
    \caption{The selection of $r$.}
  \label{tab:table5}%
\end{table}%

\subsubsection{The impact of $g$}

In Shuffle Block, the operator "channel shuffle" is to shuffle and fuse the $g$ groups of feature maps, thus the larger $g$ is, the better the features are fused, and better the performance is. In our model, the least number of channels is 32, so $g$ can only be one of the factors of 32. We set $g$ to 32, 16, and 8, respectively, and other parameters keep the same. As shown in Table \ref{tab:table6}, the larger $g$ is, the faster the model runs, and 32 is the optimal value in speed and performance. Additionally, the processing time of stage one is only $2.6\%$ of the processing time of FBP, and the total time increment is neglectable, which shows that our model has remarkable real-time performance.

\begin{table}[htbp]
  \centering
    \begin{tabular}{ccccc}
    \hline
    $g$     &
    \tabincell{c}{Radon\\PSNR/SSIM} &
    \tabincell{c}{Image\\PSNR/SSIM} &
    \tabincell{c}{Time of\\Stage\\One (ms)} &
    \tabincell{c}{Time\\in FBP\\(ms)} \\
    \hline
    \bm{$32$} & \textbf{58.849/0.9988} & \textbf{30.929/0.917} & \textbf{7.24} & 280 \\
    $16$  & 58.021/0.9987 & 30.891/0.915 & 7.29 & 280 \\
    $8$   & 57.711/0.9987 & 30.864/0.914 & 7.40 & 280 \\
    \hline
    \end{tabular}%
    \caption{The seletion of $g$.}
  \label{tab:table6}%
\end{table}%

\subsubsection{The effect of dual-domain pipeline}

\begin{figure}[t]
\centerline{\includegraphics[width=10cm]{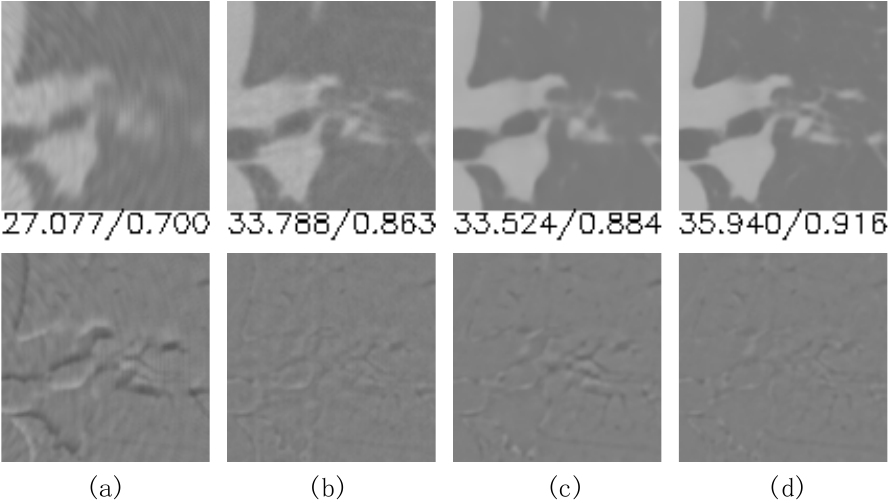}}
\caption{The reconstruction results of different repairing procedure (45 projection views). The numbers under each picture of the first row are PSNR/SSIM on this area. (a) Without any repairing. (b) One-step Radon data repairing. (c) One-step image repairing. (d) Dual-domain repairing.}
\label{sec:fig5}
\end{figure}

To verify the superiority of our dual-domain pipeline, we try to apply CAGAN directly to reconstructed images with streak artifacts. We compare the effects of one-step repairing in the Radon domain, one-step repairing in the image domain, and dual-domain repairing pipeline, as shown in Table \ref{tab:table7} and Figure \ref{sec:fig5}.

\begin{table}[htbp]
  \centering
    \begin{tabular}{cc}
    \hline
    Methods & Image PSNR/SSIM  \\
    \hline
    No Repairing + FBP & 28.722/0.833  \\
    Radon Domain Repairing + FBP & 30.929/0.917  \\
    FBP + Image Repairing & 39.620/0.955  \\
    \textbf{Dual-domain Pipeline} & \textbf{41.975/0.963}  \\
    \hline
    \end{tabular}%
    \caption{The restoration results in different stages.}
  \label{tab:table7}%
\end{table}%

Note that the evaluation metrics may lose consistency on the whole image pairs and patch pairs due to the different $data\_range$ and image structures. One-step Radon repairing is much worse than one-step image repairing on complete images, but when it comes to the case of patches, the metrics are not left behind obviously. According to both the metrics and the subjective image quality, the dual-domain pipeline performs best naturally, for the data completion in the first stage lays a good foundation, and the second stage recovers the textures and details effectively.

\begin{figure}[t]
\centerline{\includegraphics[width=15cm]{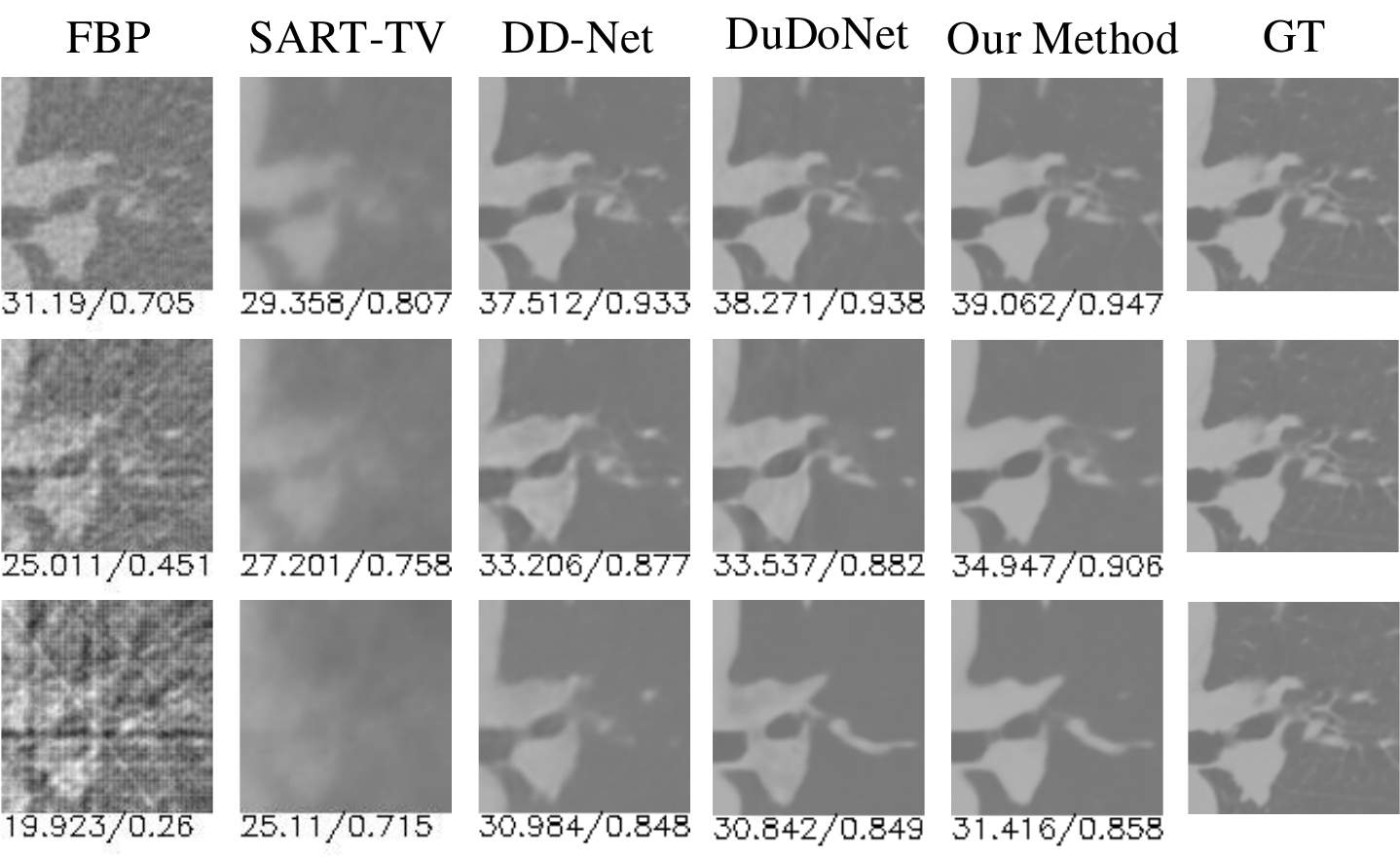}}
\caption{The zoomed region marked by the yellow boxes in Figure \ref{sec:fig7}. From the first row to the third row are 45, 20, and 10 projection views, respectively. The numbers under each picture are PSNR/SSIM in this area.}
\label{sec:fig8}
\end{figure}

\subsection{Methods Comparison}

\begin{figure*}[htbp]
\centerline{\includegraphics[width=\linewidth]{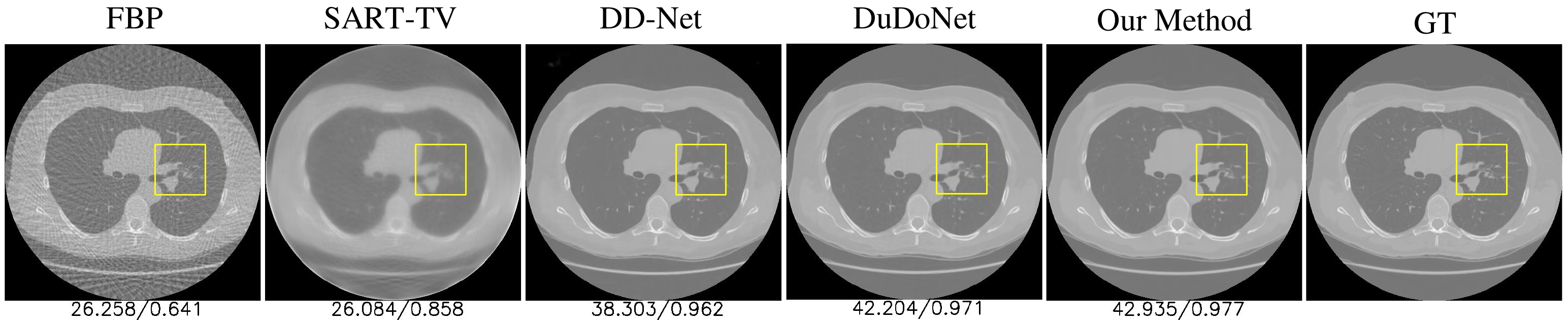}}
\caption{Different reconstruction methods on the Radon data of 45 projection views. The numbers are PSNR/SSIM on the whole picture.}
\label{sec:fig7}
\end{figure*}

We compare our algorithm with traditional analytical reconstruction method FBP \cite{katsevich2002theoretically}, iterative method SART-TV \cite{Sidky2009Accurate}, image repairing method DD-Net \cite{Zhang2018A}, and dual-domain method DuDoNet \cite{lin2019dudonet}. The parameters of SART are set as: $iters = 300$, $tv\_w = 0.1$, $tv\_maxit = 1000$, $tv\_eps = 4e-5$, $relaxation = 0.2$, respectively. DD-Net and DuDonet are retrained on our dataset, and the hyperparameters is set as same as the original paper. The results are presented in Table \ref{tab:table8}, Figure \ref{sec:fig7} and Figure \ref{sec:fig8}.

\begin{table*}[htbp]
  \centering
    \begin{tabular}{lcccc}
    \hline
        {}  & \multicolumn{3}{c}{PSNR/SSIM} & {Parameters(M)}\\
    \cmidrule{2-4}
        {}  & 45 views & 20 views & 10 views \\
    \hline
    FBP \cite{katsevich2002theoretically}  & 27.531/0.659 & 23.237/0.477 & 19.252/0.360 & - \\
    SART-TV \cite{Sidky2009Accurate} & 27.768/0.859  & 27.003/0.835 & 25.623/0.824 & - \\
    DD-Net \cite{Zhang2018A} & 40.797/0.958 & 36.841/0.935  & 33.810/0.916 & \textbf{0.3} \\
    DuDoNet \cite{lin2019dudonet} & 41.283/0.958 & 37.253/0.933 & 34.198/0.917 & 62.06 \\
    \textbf{Our Method} & \textbf{41.975/0.963} & \textbf{37.714/0.942} & \textbf{34.546/0.919} & {4.04} \\
    \hline
    \end{tabular}%
    \caption{The comparison of our method with other algorithms.}
  \label{tab:table8}%
\end{table*}%

Experiments indicate that our algorithm not only exceeds traditional reconstruction algorithms but also outperforms DD-Net and DuDoNet, achieving outstanding robustness and generalization in different sparsity of 45, 20, and 10 projection views. Especially in the case of 10 projection views, the quality of images reconstructed by FBP and SART-TV is low and unacceptable. Deep learning methods, including DD-Net, DuDoNet, and our proposed framework, achieve acceptable results, but it can be seen in Figure \ref{sec:fig8} that our results are the clearest and closest to the ground truth in the textures. Our algorithm produces images with sharp edges and without artifacts, only lost some details inevitably. It is worth mentioning that the number of parameters of our method is just $6.5\%$ of the dual-domain algorithm DuDoNet, while the performance of our pipeline shows superiority.

\section{Conclusion}

In this work, we designed the CAGAN, a lightweight adversarial auto-encoder that preserves the spatial information of images and focuses on the textures. Furthermore, we applied CAGAN into a dual-domain parallel beam CT reconstruction pipeline for sparse-view data, recovering the details of CT images. As a result, we succeed in acquiring precise, detailed, and artifact-free sparse-view CT images through our pipeline.

Moving forward, the architecture of CAGAN might be able to be applied to other image-to-image tasks, such as image enhancing, and semantic segmentation, which is worthy of our further research. Also, extreme sparsity of projections would lead to much loss of information and cause severe artifacts beyond any repairing algorithms. Therefore, given the specific image quality, the research about the lower limit of CT scanning sparsity has aroused our interest.


\bibliographystyle{unsrt}
\bibliography{references}

\end{document}